\documentclass[12pt, preprint, authoryear]{elsarticle}

\usepackage{lineno,hyperref}
\modulolinenumbers[5]
\hypersetup{
	colorlinks   = true, %Colours links instead of ugly boxes
}

\usepackage{amsmath}
\usepackage{amsthm}

\newtheoremstyle{new}{12pt}{12pt}{\itshape}{}{\bfseries}{.}{1em}{}
\theoremstyle{new}
\newtheorem{proposition}{Proposition}

\newtheorem{definition}{Definition}
\newtheorem{example}{Example}

\usepackage{bm}
\usepackage{amsfonts}
\usepackage{subfig}
\usepackage{graphicx}
\usepackage{dsfont}

\usepackage{nameref}
\usepackage{color}
\definecolor{TUMblue}{cmyk}{1, .54, .04, .19}

\newcommand{\ind}{\mathds{1}}
\newcommand{\R}{\mathbb{R}}
\newcommand{\N}{\mathbb{N}}
\newcommand{\Z}{\mathbb{Z}}

\renewcommand{\Pr}{\mathrm{Pr}}

\newcommand{\wh}{\widehat}

\usepackage{setspace}
\begin{document}
	% title page
	\title{A generic approach to nonparametric function estimation with mixed data}
%\footnote{Department of Mathematics, Technische Universit{\"a}t M{\"u}nchen, Boltzmanstra{\ss}e 3,  85748 Garching, Germany \href{mailto:mail@tnagler.com}{mail@tnagler.com}}}

\author{Thomas Nagler}
\address{Department of Mathematics, Technical University of Munich, Boltzmanstra{\ss}e 3,  85748 Garching, Germany \href{mailto:mail@tnagler.com}{mail@tnagler.com}}

\begin{abstract} 
\noindent Most nonparametric function estimators can only handle continuous data. We show that making discrete variables continuous by adding noise is justified under suitable conditions on the noise distribution. This principle is widely applicable, including density and regression function estimation.
\end{abstract}


\maketitle

	% content
	\section{Introduction}

In applications of statistics, data containing discrete variables are omnipresent. An online retailer records information on how many purchases a customer made in the past. Social scientists typically use discrete scales on which study participants rate their satisfaction, attitude, or feelings. Another common example is where data describe unordered categories, like gender or business sectors.

Suppose that $(\bm Z, \bm X)$ is a random vector with discrete component $\bm Z \in \Z^p$ and continuous component $\bm X \in \R^q$. This includes the cases $p \ge 1$, $q = 0$ (all variables are discrete) and $p = 0$, $q\ge 1$ (all variables are continuous). We consider problems where one aims at estimating a functional $T$ of the density/probability mass function $f_{\bm{Z}, \bm X}$ based on observations $({\bm{ Z}_i, \bm X_i})$, $i = 1, \dots, n$.  This formulation is general enough to include many common problems in nonparametric function estimation, in particular: density estimation, regression, and classification.

Some nonparametric estimation techniques have been specifically designed to allow for mixed continuous and discrete data \citep{Ahmad94, Li03, Hall83, Efromovich11}, but the number is small and the more sophisticated methods are often developed in a purely continuous framework. Examples are local polynomial methods \citep{fan1996local, Loader99} or copula-based estimators \citep[e.g.,][]{Otneim16,Nagler16,Kauermann14}. These methods are no longer consistent when applied to mixed data types.

There is a popular trick among practitioners to get an approximate answer nevertheless: just make the data continuous by adding noise to each discrete variable. This trick is sometimes called \emph{jittering} or \emph{adding jitter}.  Examples where it has been successfully applied are: avoiding overplotting in data visualization \citep{Few08}, adding intentional bias to complex machine learning models \citep{Zur04}, deriving theoretical properties of concordance measures \citep{Denuit05}, or nonparametric copula estimation for mixed data \citep{genest2017asymptotic}. An example of its misuse was pointed out by \citet{Nikol13} in the context of parametric copula models. Generally, the trick lacks theoretical justification because it can introduce bias. But we shall see that this issue is resolved under a suitable choice of noise distribution.

This letter aims to formalize this somewhat ``dirty'' trick and to provide a starting point for a more nuanced investigation of its properties. Some open questions and partial answers will be given at the end.

    \section{Jittering mixed data} \label{CE}

\subsection{Preliminaries and notation}

We assume throughout that all random variables live in a space with a natural concept of ordering. Unordered categorical variables can always be coded into a set of binary dummy variables (for which $0 < 1$ gives a natural ordering). We further assume without loss of generality that any discrete random variable, say $Z$, is supported on a set $\Omega_Z \subseteq \Z$. For any continuous random vector $\bm X$, we write $f_{\bm X}$ for its joint density. In case $\bm Z$ is a discrete random vector, $f_{\bm Z}$ denotes its density with respect to the counting measure, i.e., $f_{\bm Z}(\bm z) = \Pr(\bm Z = \bm z)$. A random vector with mixed types will be partitioned into  $(\bm Z, \bm X) \in \Z^p \times \R^q$. Then $f_{\bm Z, \bm X}$ is the density with respect to the product of the counting and Lebesgue measures,
\begin{align*}
    f_{\bm Z, \bm X}(\bm z, \bm x) =  \frac{\partial^{q}}{\partial x_{1} \cdots \partial x_{q}} \Pr\bigl(\bm{Z} = \bm z, \bm{X} \le \bm x\bigr).
\end{align*}

%\begin{remark}
%The noise density $\eta$ may exhibit jumps at a finite number of points in $\R$. An example of such a density is the uniform density on $(-0.5, 0.5)$, which jumps at $-0.5$ and $0.5$. 
%\end{remark}

%\begin{figure}
%	\centering
%    \subfloat[original data]{
%    \includegraphics[width = 0.315\textwidth]{fig/cx_ill_1} 
%     \label{cx_ill_1}
%     } 
%     \hspace{2cm}
%     \subfloat[jittered data]{
%    \includegraphics[width = 0.315\textwidth]{fig/cx_ill_2}
%    \label{cx_ill_2}
%    }  
%    \caption{Adding jitter to data with mixed types. The discrete component $Z$ is convoluted with $E \sim \mathrm{Uniform}(-0.5, 0.5)$.}
%    \label{cx_ill}
%\end{figure}

%\begin{example}
%Figure \ref{cx_ill_1} shows simulated data of a random vector $(Z, X) \in \Z \times \R$. The discrete component $Z$ takes values in the set $\{0, 1, 2, 3, 4\}$; the continuous component $X$ is supported on $\R_{+}$. The jittered version of $(Z, X)$ is constructed by adding $\mathrm{Uniform}(-0.5, 0.5)$ noise to the discrete component $Z$. The jittered data is shown in Fig.~\ref{cx_ill_2}. They take values in $(-0.5, 4.5) \times \R_{+}$ and represent observations from a purely continuous random vector.
%\end{example}

\subsection{The density of a jittered random vector}

The jittered version of a random vector is defined by adding noise to all discrete variables. 

\begin{definition} \label{ce:ce_def}
Let $\eta$ be a bounded density function that is continuous almost everywhere on $\R$. The jittered version of the random vector $(\bm Z, \bm X) \in \Z^{p} \times \R^{q}$ is defined as $(\bm Z + \bm \epsilon, \bm X)$, where $\bm \epsilon \in \R^p$ is independent of $(\bm Z, \bm X)$.
\end{definition}

Provided that $f_{\bm Z, \bm X}$ exists, the density of the jittered vector $(\bm Z + \bm \epsilon, \bm X)$ is simply the discrete-continuous convolution of $f_{\bm Z, \bm X}$ and the noise density $f_{\bm \epsilon}$:
\begin{align*}
f_{\bm Z + \bm \epsilon, \bm X}(\bm z, \bm x) = \sum_{\bm z^\prime \in \Z^p} f_{\bm Z, \bm X}(\bm z^\prime, \bm x) f_{\bm \epsilon}(\bm z - \bm z^\prime), \quad \mbox{for almost all } (\bm z, \bm x) \in \R^{p + q}.
\end{align*}
We observe a close relationship between the densities $f_{\bm Z + \bm \epsilon, \bm  X}$ and $f_{\bm Z, \bm X}$. If we know $f_{\bm Z, \bm X}$ at all values $(\bm z^\prime, \bm x) \in \Z^p \times \R^q$, we can immediately compute $f_{\bm Z + \bm \epsilon, \bm X}$ at all values $(\bm z, \bm x) \in \R^{p \times q}$. The other direction is more interesting for our purposes: can we recover  $f_{\bm Z, \bm X}$  from known values of $f_{\bm Z + \bm \epsilon, \bm X}$? In general, this poses a  rather challenging deconvolution problem. But we can make things easier by a suitable choice of noise density $\eta$. In fact, there is a large class of noise densities densities for which no deconvolution is necessary and $f_{\bm Z, \bm X}$ and $f_{\bm Z + \bm \epsilon, \bm X}$ coincide on $\Z^p \times \R^q$.

\begin{proposition} \label{ce:Bdens_prop}
It holds 
\begin{align} \label{eq:equality}
 f_{\bm Z + \bm \epsilon, \bm X}(\bm z, \bm x) =  f_{\bm Z, \bm X}(\bm z, \bm x)
\end{align}
for any joint density $f_{\bm Z, \bm X}$ and all $(\bm z, \bm x) \in \Z^p \times \R^q$, if and only if the following two conditions are satisfied:
\begin{enumerate}
\item $f_{\bm \epsilon}(\bm 0) = 1$,
\item there exists $\gamma_2 \in (0, 1)$ such that $f_{\bm \epsilon}(\bm x) = 0$ for all $\bm x \in \R^p \setminus [-\gamma_2, \gamma_2]^p$. 
\end{enumerate}
\end{proposition}
A simple, but powerful implication is that we can estimate the discrete-continuous density $f_{\bm Z, \bm X}$ by estimating the purely continuous density $f_{\bm Z + \bm \epsilon, \bm X}$.

\subsection{A convenient class of noise distributions} \label{ce:noise}

In the following we give a particularly convenient class of noise densities.

\begin{definition} \label{ce:BA_def}
We say that $f_{\bm \epsilon} \in \mathcal{E}_{\gamma_1, \gamma_2}$ for some $0 < \gamma_1 \le 0.5 \le \gamma_2 < 1$, if $f_{\bm \epsilon}(\bm x) = \prod_{j=1}^p \eta(x_p)$ for all $\bm x \in \R^p$, where $\eta$ is an absolutely continuous probability density function, $\eta(x) = 1$ for all $x \in [-\gamma_1, \gamma_1]$, and $\eta(x) = 0$  for all $x \in \R \setminus (-\gamma_2, \gamma_2)$.
\end{definition}

The class $\mathcal{E}_{\gamma_1, \gamma_2}$ satisfies \eqref{eq:equality}, but adds two more restrictions to the conditions given in Proposition~\ref{ce:Bdens_prop}: (i) the random noise is componentwise independent, (ii) it is constant in a neighborhood of zero. The first restriction is made purely for convenience and will be discussed further in Section \ref{sec:issues}. The second ensures that the derivatives of $f_{\bm Z + \bm \epsilon, \bm X}(\bm z, \bm x)$ with respect to $\bm z$ vanish for all $(\bm z, \bm x) \in \Z^p \times \R^q$. This property is particularly useful in nonparametric density estimation, since an estimators' bias is usually proportional to derivatives of the target density.

\begin{proposition} \label{ce:BCdens_prop}
If  $f_{\bm \epsilon} \in \mathcal{E}_{\gamma_1, \gamma_2}$, $(\bm z, \bm x) \in \Z^{p} \times \R^q$, and $\bm m \in \N^p$ such that $\sum_{k = 1}^p m_k = \overline m$, then

\begin{align*}
	\frac{\partial^{\overline m} f_{\bm Z + \bm \epsilon, \bm X}(\bm z, \bm x)}{\partial z_1^{m_1} \cdots \partial z_p^{m_p}} = 0.
\end{align*}
\end{proposition}

\begin{figure}[t]
\centering
\subfloat[original density]{
\includegraphics[width = 0.3\textwidth]{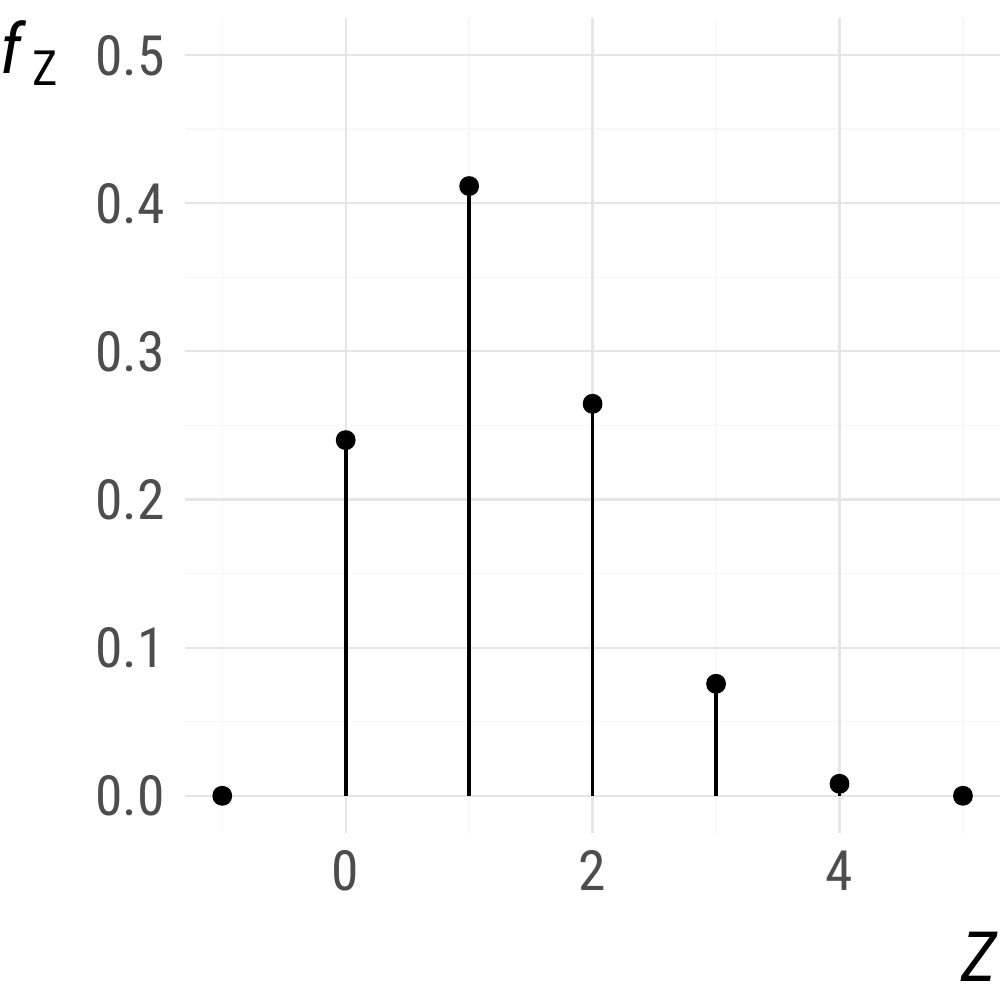}
\label{ce:unif1_fig}
} 
\subfloat[noise density]{
\includegraphics[width = 0.3\textwidth]{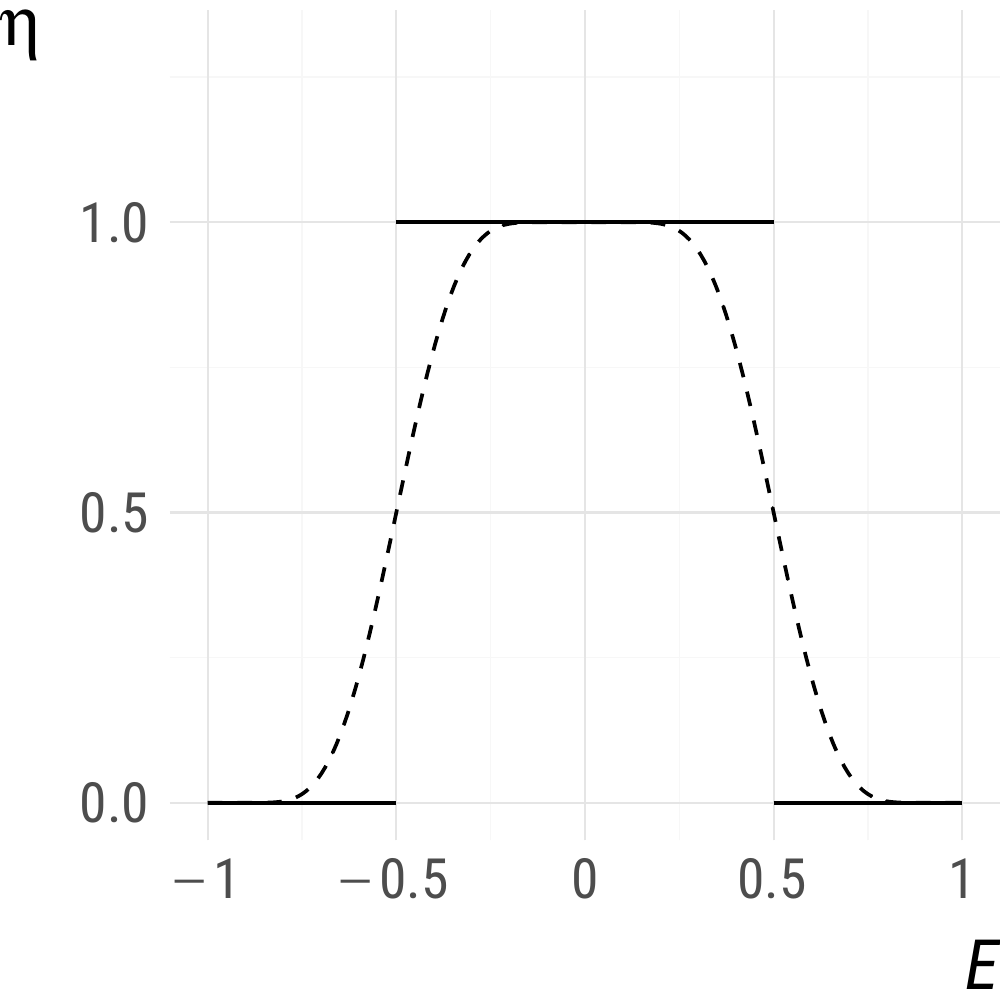}
\label{ce:unif2_fig}
} 
\subfloat[convoluted density]{
\includegraphics[width = 0.3\textwidth]{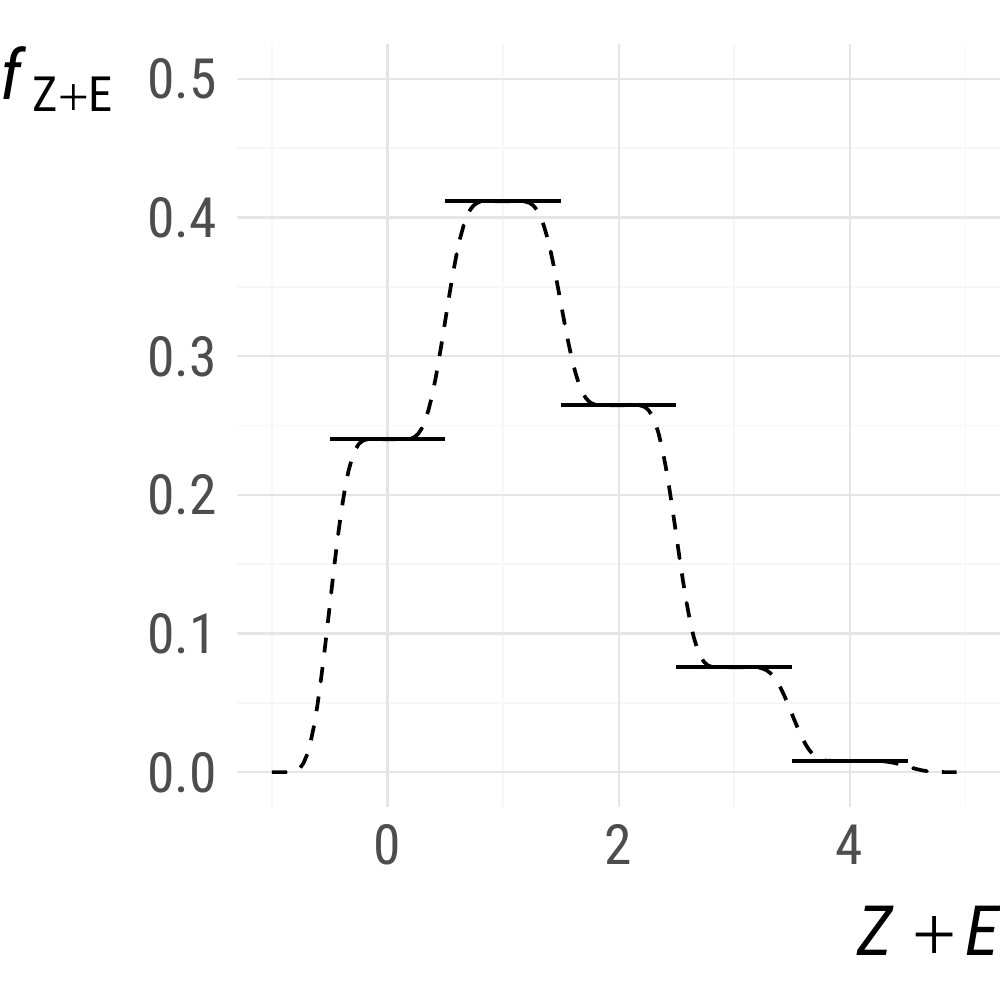}
\label{ce:unif3_fig}
} 
\caption{Jittering of a density function: (a) $\mathrm{Binomial}(4, 0.3)$ density; (b) two noise densities from the family $f_{U_{\theta, \nu}}$, see Example~\ref{ce:usb_ex}; (c) the convolution of the densities in (a) and (b).}
\label{ce:unif_fig}
\end{figure}

\begin{example} \label{ce:usb_ex}
Let $\nu \in \N$ and $0 \le \theta < 1$. Set $U_{\theta, \nu} = U + \theta (B_\nu - 0.5)$ where $U \sim \mathcal{U}(-0.5, 0.5)$ and $B_\nu \sim \mathrm{Beta}(\nu, \nu)$.  
The density of $U_{\theta, \nu}$ can be calculated as

\begin{align*}
f_{U_{\theta, \nu}}(x) = \begin{cases}
\ind(\vert x \vert < 0.5), & \theta = 0, \\
F_{B_\nu}\bigl\{(x + 0.5) / \theta + 0.5\bigr\} - F_{B_\nu}\bigl\{(x - 0.5) / \theta + 0.5\bigr\}, & \theta > 0.
\end{cases}
\end{align*}
It is easy to check that $f_{U_{\theta, \nu}} \in \mathcal{E}_{(1 - \theta)/2, (1 + \theta)/2}$ and that $f_{U_{\theta, \nu}}$ is $\nu - 1$ times continuously differentiable everywhere on $\R$. Hence, if $f_{\bm Z, \bm X}(\bm z, \bm x)$ is $m$ times continuously differentiable in $\bm x$ for all $(\bm z, \bm x) \in \Z^p \times \R^q$, $f_{\bm Z + \bm \epsilon, \bm X}$ is $\min\{\nu - 1, m\}$ times continuously differentiable everywhere on $\R^{p + q}$. Also, $f_{\bm Z + \bm \epsilon, \bm X}$ coincides with $f_{\bm Z, \bm X}$ everywhere on $\Z^p \times \R^q$. This is illustrated in Fig.~\ref{ce:unif_fig} for $\theta = 0$ (the uniform distribution on $(-0.5,0.5)$, solid), as well as  $\theta = 0.8$ and $\nu = 5$ (dashed). \qed
\end{example}

    \section{Nonparametric function estimation via jittering} \label{npest}

\subsection{Jittering estimators}

Suppose we want to estimate a functional $T$ of $f_{\bm Z, \bm X}$, where $(\bm Z, \bm X) \in \Z^p \times \R^q$. Let $(\bm Z_1, \bm X_1)$, \dots, $(\bm Z_n, \bm X_n)$ be a stationary sequence of random vectors having the same distribution as $(\bm Z, \bm X)$.  Let further $\bm \epsilon_{i}$, $i = 1, \dots, n$, be independent and identically distributed vectors that have the same distribution as $\bm \epsilon$ (as in Definition \ref{ce:ce_def}) and are independent of $(\bm Z_1, \bm X_1), \dots, (\bm Z_n, \bm X_n)$. 

\begin{definition} \label{npest:cce_def}
An estimator $\wh \tau_n$ of $T(f_{\bm Z, \bm X})$ is called jittering estimator if it is a measurable function of the jittered data, i.e., $\wh \tau_n = \wh \tau_n(\bm Z_1 + \bm \epsilon_1, \dots, \bm X_n)$.
\end{definition}

Jittering estimators are extremely easy to implement: all one needs is a way to generate random noise and an estimator that works for continuous data. The following two examples introduce jittering analogues of popular estimators that, in their original version, are only applicable to continuous data.
\begin{example}[Kernel density estimation] \label{ex:kde}
The jittering kernel density estimator of $f_{\bm Z, \bm X}$ is
\begin{align*}
\widetilde f(\bm z, \bm x) = \frac{1}{nb_n} \sum_{i = 1}^n K\biggl\{\frac{(\bm Z_i + \bm \epsilon_i, \bm X_i) - (\bm z, \bm x)}{b_n}\biggr\}.
\end{align*}
where $b_n > 0$ and $K$ is a symmetric, multivariate density function. The classical kernel density estimator of \citet{Parzen62} and \citet{Rosenblatt56} is recovered when $\bm \epsilon_i = 0$ for all $i = 1, \dots, n$.
\end{example}

\begin{example}[Local linear regression]
The jittering local linear regression estimator $\wh m$ of $E(X_1 \mid \bm Z = \bm z, \bm X_{-1} = \bm x_{-1})$ is 
\begin{align*}
\underset{(\wh m, \bm \beta) \in \R^{p + q - 1}}{\arg\min} \sum_{i = 1}^n \bigl\{ \wh m - X_{i, 1} - \bm \beta^\top (\bm Z_i + \bm \epsilon_i, \bm X_{i, -1}) +  \bm \beta^\top (\bm z, \bm x_{-1}) \bigr\},
\end{align*}
where $b_n$ and $K$ are as in Example \ref{ex:kde}. With $\bm \epsilon_i = 0$ for all $i = 1, \dots, n$, we recover the classical local linear regression estimator \citep[e.g.,][]{fan1996local}.
\end{example}

\subsection{Applications: estimating a regression function}

Now suppose that there is another functional $T^*$ such that $T(f_{\bm Z, \bm X}) = T^*(f_{\bm Z + \bm \epsilon, \bm X})$. We shall call $T^*$ the \emph{jittering equivalent} of $T$. Now if $\wh \tau$ is an estimator of $T^*(f_{\bm Z + \bm \epsilon, \bm X})$, then it is also an estimator of $T(f_{\bm Z, \bm X})$. This means that we can use any estimator that works in a purely continuous setting to estimate the target functional $T(f_{\bm Z, \bm X})$, even though $f_{\bm Z, \bm X}$ is the density of a mixed data model. An example for such a situation is density estimation where $T(f_{\bm Z, \bm X}) = f_{\bm Z, \bm X}(\bm z, \bm x)$ and $T^* = T$ (see Proposition~\ref{ce:Bdens_prop}). But the setup is much more general and covers most common regression problems, as the following examples show.

\begin{example}[Mean regression, continuous response] \label{ex:mreg}
 The conditional mean $\mathrm{E}(X_1 \mid \bm Z = \bm z, \bm X_{-1} = \bm x_{-1})$ can be expressed as
\begin{align*}
T_{m, c}(f_{\bm Z, \bm X}) =
 \biggl\{ \int_{\R} f_{\bm Z, \bm X}(\bm z, \bm x) dx_1 \biggr\}^{-1} \int_{\R} x_1 f_{\bm Z, \bm X}(\bm z, \bm x) dx_1.
\end{align*}
The jittering equivalent is $T^*_{m,c} = T_{m,c}$. The discrete response case is analogous.
\end{example}

\begin{example}[Distribution regression, discrete response] \label{ex:dreg}
 The conditional distribution function $\Pr(Z_1 \le \bar z_1  \mid \bm Z_{-1} = \bm z_{-1}, \bm X = \bm x)$ can be expressed as
\begin{align*}
T_{p,d}(f_{\bm Z, \bm X}) =
\biggl\{ \sum_{z_1 \in \Z} f_{\bm Z, \bm X}(\bm z, \bm x) \biggr\}^{-1} \sum_{z_1 = - \infty}^{\bar z_1}  f_{ \bm Z, \bm X}(\bm z, \bm x).
\end{align*}
The jittering equivalent is 
\begin{align*}
T^*_{p,d} =\biggl\{ \int_{\R} f_{\bm Z+ \bm \epsilon, \bm X }(\bm z, \bm x) dz_1 \biggr\}^{-1} \int_{-\infty}^{\bar z_1} f_{\bm Z+ \bm \epsilon, \bm X}(\bm z, \bm x) dz_1 + \frac{f_{\bm Z + \bm \epsilon, \bm X}(\bar z_1, \bm z_{-1}, \bm x) }{2 \int_{\R} f_{\bm Z + \bm \epsilon, \bm X}(\bm z, \bm x) dz_1}.
\end{align*}
The continuous response case is similar, but does not require a correction term as in the previous display.

\end{example}

\begin{example}[Quantile regression] \label{ex:qreg}
For $\alpha \in [0, 1]$, the conditional quantile function corresponding to $\Pr(Z_1 \le \cdot \mid \bm Z_{-1} = z_{-1}, \bm X = \bm x)$ can be expressed as $
T_{q,d}(f_{\bm Z, \bm X}) = \inf\bigl\{\bar z_1 \in \R \colon  T_{p,d}(f_{\bm Z, \bm X})  \ge \alpha  \bigr\},$
where $T_{p,d}$ is as in Example \ref{ex:dreg}. The jittering equivalent is
$T_{q,d}^*(f_{\bm Z, \bm X}) = \inf\bigl\{\bar z_1 \in \R \colon  T_{p,d}^*(f_{\bm Z, \bm X})  \ge \alpha  \bigr\}.$
The continuous response case is analogous.
\end{example}

\subsection{Asymptotic properties}

A convenient fact about jittering estimators is that asymptotic properties for estimating $T^*(f_{\bm Z + \bm \epsilon, \bm X})$ directly translate into properties for estimating $T(f_{\bm Z, \bm X})$. The following result is trivial, but important enough to be stated formally.

\begin{proposition}
Let $T$ and $T^*$ be two functionals such that $T(f_{\bm Z, \bm X}) = T^*(f_{\bm Z + \bm \epsilon, \bm X})$. If for some sequence $r_n \to 0$ and random variable $W$, $r_n^{-1}\{\wh \tau - T^*(f_{\bm Z + \bm \epsilon, \bm X})\} \to W$ almost surely, in probability, or in distribution, then also $r_n^{-1}\{\wh \tau - T(f_{\bm Z, \bm X})\} \to W$ almost surely, in probability, or in distribution.
\end{proposition}

In particular, any (strongly) consistent estimator of $T^*(f_{\bm Z + \bm \epsilon, \bm X})$ is at the same time a (strongly) consistent estimator of $T(f_{\bm Z, \bm X})$. Even better: since we can choose the noise distribution $\eta$ we gain some control over the local behavior of the jittered density $f_{\bm Z + \bm \epsilon, \bm X}$. If $T^*$ is sufficiently well-behaved, this allows us to control the local behavior of the estimation target $T^*(f_{\bm Z + \bm \epsilon, \bm X})$, too. For example, the form of the regression functionals and Proposition \ref{ce:BCdens_prop} imply that all derivatives of $T^*(f_{\bm Z + \bm \epsilon, \bm X})$ w.r.t.\ $\bm z$ vanish in a $\gamma_1$-neighborhood of $\bm z \in \Z^p$. This allows to estimate regression functionals without bias for the discrete part and, thus, to improve the convergence rates of the estimator $\wh \tau_n$; see \citet{nagler2017asymptotic} for an in-depth analysis of the jittering kernel density estimator.

%However, asymptotic properties derived this way can be suboptimal. For example, the density $f_Z$ of a discrete variable $Z \in \Z$ can be estimated consistently by the sample frequency $f_n(z) = n^{-1} \sum_{i = 1}^n \ind(Z_i = z)$, for which $f_Z(z) - f_n(z) = O_p(n^{-1/2})$. In contrast, the best possible convergence rate for estimating a twice continuously differentiable density $f_X$ of a continuous variable $X \in \R$ is $n^{-1/3}$ \citep[see,][]{Stone80}. This does not mean that jittering estimators are generally inefficient. But the regularity conditions imposed in continuous models are usually stronger than what is needed in the mixed data setting. This is illustrated in the following example.
%
%\begin{example}[Kernel density estimation for a discrete variable]
%The classical kernel density estimator with uniform kernel is defined as $\wh k(z) = (n b_n)^{-1} \sum_{i=1}^n\ind(\vert z - Z_i \vert \le b_n)$. If $Z$ is continuous, $b_n \sim n^{-1/3}$, and $f_Z$ is twice continuously differentiable at $z$, it holds $\wh k(z) - f_Z(z) = O_p(n^{-1/3})$ \citep[e.g.,][]{Parzen62}. We will show that this rate can be improved for the jittering version of $\wh k$. Let $Z \in \Z$ and define $\widetilde k$ as the jittering version of $\wh k$, i.e., $\widetilde k(z) = (n b_n)^{-1} \sum_{i=}^n\ind(\vert z - Z_i - E_i) \vert \le b_n)$. Assuming $\eta(x) = \ind(\vert x \vert \le 0.5)$ and $0 < b_n \le 0.5$, it holds $\widetilde k \equiv f_n$ and, thus, $\widetilde k(z) - f_Z(z) = O_p(n^{-1/2})$.
%\end{example}

    %\input{5-application}
	\section{Discussion}

\subsection{Benefits}

The most obvious benefit of jittering estimators is convenience. For their implementation, all one needs is an estimator that works in the continuous setting and a way to simulate random noise. This is easily achieved in modern statistical software. At second glance, the method opens many possibilities to extend existing estimators to the mixed data setting. This is increasingly useful with increasing complexity of the estimators. In many cases, there is otherwise no straightforward way to adapt an estimator to mixed data. 

A less obvious benefit arises for studying general properties of a nonparametric function estimation problem. In the continuous setting, asymptotic arguments are often easier and well-established. For example, jittering arguments make it straightforward to derive minimax-optimal rates of convergence in nonparametric mixed data models; see \citet{nagler2017asymptotic} in the case of density estimation.

\subsection{Issues and open questions} \label{sec:issues}

\subsubsection*{Curse of dimensionality}

A key issue for nonparametric estimators is the \emph{curse of dimensionality}. In a continuous setting, the speed of convergence decreases exponentially in the dimension. For example, the classical convergence rate for estimating a $d$-dimensional continuous density is $n^{-2/(4 + d)}$. A discrete density on the other hand can always be estimated with $n^{-1/2}$ rate, so there is no curse of dimensionality. It is not obvious, which regime jittering estimators fall into, since a discrete density is estimated by exchanging it with a continuous surrogate. 

Unfortunately, this question has no general answer and depends on the estimators' characteristics.  The main criterion is how ``local'' the estimator operates; or more specifically, if the estimator is only affected by data in a compact neighborhood. For example, B-spline methods and kernel estimators with a compact kernel function will usually fall into the discrete regime, whereas Bernstein polynomials and kernel estimators with unbounded kernels fall into the continuous one. But we should stress that such considerations are only asymptotic and the behavior on finite samples will likely fall somewhere in between. 

\subsubsection*{Efficiency}

Typically, adding noise brings about some unnecessary variance. The magnitude of this effect depends on the characteristics of the estimator. Generally, this additional variance can be reduced by averaging estimates over multiple independent jitters \citep[cf.,][]{genest2017asymptotic}. In specific cases, a jittering estimator can be inherently efficient, with no need for averaging \citep[see,][Section 4.1]{nagler2017asymptotic}.

\subsubsection*{Choice of noise distribution}

When using the jittering technique, an immediate question is which noise distribution to choose. The necessary conditions given in Proposition 1 are fairly broad and allow for a variety of noise distributions. 

A referee asked whether it would be possible to preserve some dependence characteristics of the data. Unfortunately, dependence between discrete variables and its connection to the continuous counterpart is a highly subtle issue. One such subtlety is that there is no density when continuous variables are perfectly dependent, but the probability mass function for perfectly dependent variables exists. \citet{Genest2007} address many other interesting issues. The article also provides some arguments for using independent noise, because it is the only way to preserve the equality between probabilistic and analytical definitions of some margin-free dependence measures like Kendall's $\tau$ and Spearman's $\rho$ (their equation 7) or tie-corrected versions (p.\ 495).

In any case, one should understand jittering as an estimation technique rather than a modeling technique. Interpreting the jittered model independently of the ``true'' one is unlikely to be beneficial. The letter's only criterion for validity of jittering was consistency of estimators.  But we should expect that a data-driven choice of noise distribution would improve estimators' accuracy. A closer examination of the noise distribution's effect will be a promising path for future research. 

\subsubsection*{Restriction to nonparametric techniques}

Finally, we should warn that this methodology is only valid for nonparametric estimators. Usually, the shape of functionals of the jittered density can not be captured by parametric models, leading to estimators that are inconsistent.

\subsection*{Acknowledgements}

This work was partially supported by the German Research Foundation (DFG grant CZ 86/5-1). The author thanks two anonymous referees for raising many interesting points that greatly improved the comprehensiveness of this contribution.

%\subsection*{Supplementary material}

% \begin{itemize}
% \item \href{https://github.com/tnagler/cctools}{https://github.com/tnagler/cctools}: an R package providing tools for jittering.
% \item \href{https://github.com/tnagler/cctools}{https://github.com/tnagler/jdify}: an R package providing functionality for joint density classification.
% \item \href{https://gist.github.com/tnagler/843f5c658e1139ff669d33614cc727e6}{https://gist.github.com/tnagler/843f5c658e1139ff669d33614cc727e6}: R code replicating the results from Section \ref{application}.
% \end{itemize}

	% load references
	\bibliographystyle{elsarticle-harv}
	\section*{References}
	\small
	\bibliography{references}

\begin{thebibliography}{18}
\expandafter\ifx\csname natexlab\endcsname\relax\def\natexlab#1{#1}\fi
\expandafter\ifx\csname url\endcsname\relax
  \def\url#1{\texttt{#1}}\fi
\expandafter\ifx\csname urlprefix\endcsname\relax\def\urlprefix{URL }\fi

\bibitem[{Ahmad and Cerrito(1994)}]{Ahmad94}
Ahmad, I.~A., Cerrito, P.~B., 1994. Nonparametric estimation of joint
  discrete-continuous probability densities with applications. Journal of
  Statistical Planning and Inference 41~(3), 349--364.

\bibitem[{Denuit and Lambert(2005)}]{Denuit05}
Denuit, M., Lambert, P., 2005. Constraints on concordance measures in bivariate
  discrete data. Journal of Multivariate Analysis 93~(1), 40--57.

\bibitem[{Efromovich(2011)}]{Efromovich11}
Efromovich, S., 2011. Nonparametric estimation of the anisotropic probability
  density of mixed variables. Journal of Multivariate Analysis 102~(3), 468 --
  481.

\bibitem[{Fan and Gijbels(1996)}]{fan1996local}
Fan, J., Gijbels, I., 1996. Local polynomial modelling and its applications:
  monographs on statistics and applied probability 66. Vol.~66. CRC Press.

\bibitem[{Few(2008)}]{Few08}
Few, S., 2008. Solutions to the problem of over-plotting in graphs. Visual
  Business Intelligence Newsletter.

\bibitem[{Genest and Neslehova(2007)}]{Genest2007}
Genest, C., Neslehova, J., 2007. A primer on copulas for count data. Astin
  Bulletin 37~(02), 475--515.

\bibitem[{Genest et~al.(2017)Genest, Ne{\v{s}}lehov{\'a}, and
  R{\'e}millard}]{genest2017asymptotic}
Genest, C., Ne{\v{s}}lehov{\'a}, J.~G., R{\'e}millard, B., 2017. Asymptotic
  behavior of the empirical multilinear copula process under broad conditions.
  Journal of Multivariate Analysis.

\bibitem[{Hall et~al.(1983)}]{Hall83}
Hall, P., et~al., 1983. Orthogonal series methods for both qualitative and
  quantitative data. The Annals of Statistics 11~(3), 1004--1007.

\bibitem[{Kauermann and Schellhase(2014)}]{Kauermann14}
Kauermann, G., Schellhase, C., 2014. Flexible pair-copula estimation in d-vines
  with penalized splines. Statistics and Computing 24~(6), 1081--1100.

\bibitem[{Li and Racine(2003)}]{Li03}
Li, Q., Racine, J., 2003. Nonparametric estimation of distributions with
  categorical and continuous data. Journal of Multivariate Analysis 86~(2),
  266--292.

\bibitem[{Loader(1999)}]{Loader99}
Loader, C., 1999. Local regression and likelihood. Springer New York.

\bibitem[{Nagler(2017)}]{nagler2017asymptotic}
Nagler, T., 2017. Asymptotic analysis of the jittering kernel density
  estimator. arXiv:1705.05431.

\bibitem[{Nagler and Czado(2016)}]{Nagler16}
Nagler, T., Czado, C., 2016. Evading the curse of dimensionality in
  nonparametric density estimation with simplified vine copulas. Journal of
  Multivariate Analysis 151, 69--89.

\bibitem[{Nikoloulopoulos(2013)}]{Nikol13}
Nikoloulopoulos, A.~K., 2013. On the estimation of normal copula discrete
  regression models using the continuous extension and simulated likelihood.
  Journal of Statistical Planning and Inference 143~(11), 1923--1937.

\bibitem[{Otneim and Tj{\o}stheim(2016)}]{Otneim16}
Otneim, H., Tj{\o}stheim, D., 2016. The locally gaussian density estimator for
  multivariate data. Statistics and Computing, 1--22.

\bibitem[{Parzen(1962)}]{Parzen62}
Parzen, E., 09 1962. On estimation of a probability density function and mode.
  The Annals of Mathematical Statistics 33~(3), 1065--1076.
\newline\urlprefix\url{http://dx.doi.org/10.1214/aoms/1177704472}

\bibitem[{Rosenblatt(1956)}]{Rosenblatt56}
Rosenblatt, M., 09 1956. Remarks on some nonparametric estimates of a density
  function. The Annals of Mathematical Statistics 27~(3), 832--837.
\newline\urlprefix\url{http://dx.doi.org/10.1214/aoms/1177728190}

\bibitem[{Zur et~al.(2004)Zur, Jiang, and Metz}]{Zur04}
Zur, R., Jiang, Y., Metz, C., 2004. Comparison of two methods of adding jitter
  to artificial neural network training. International Congress Series 1268,
  886 -- 889, \{CARS\} 2004 - Computer Assisted Radiology and Surgery.
  Proceedings of the 18th International Congress and Exhibition.

\end{thebibliography}
		
\end{document}